\documentclass[longauth,letter]{aa} % for the letters
\usepackage{graphicx}
\usepackage{hyperref}
\usepackage{natbib}                 % improved management of bibliographic citations
\bibpunct{(}{)}{;}{a}{}{,}          % to follow the A&A style for the bibliography
\usepackage{amsmath}                % AMS Math
\usepackage{txfonts}
\newcommand{\rgb}{RGB\,J0152$+$017}
\newcommand{\fluxunits}{cm$^{-2}$\,s$^{-1}$}
\newcommand{\difffluxunits}{cm$^{-2}$\,s$^{-1}$\,TeV$^{-1}$}
\newcommand{\spectralindex}{$\Gamma=2.95\pm0.36_{\mathrm{stat}}\pm 0.20_{\mathrm{syst}}$}
\newcommand{\normalization}{$\Phi(1 \mathrm{TeV}) = (5.7 \pm 1.6_{\mathrm{stat}}\pm 1.1_{\mathrm{syst}})\times 10^{-13}$ \difffluxunits}
\newcommand{\spectralindexstd}{$\Gamma=3.53\pm0.60_{\mathrm{stat}}\pm 0.2_{\mathrm{syst}}$}
\newcommand{\normalizationstd}{$\Phi(1 \mathrm{TeV}) = (4.4 \pm 2.0)\times 10^{-13}$ \difffluxunits}

\begin{document}
\title{Discovery of VHE $\gamma$-rays from the high-frequency-peaked\\
 BL\,Lac object \rgb}

\author{F. Aharonian\inst{1,13}
 \and A.G.~Akhperjanian \inst{2}
 \and U.~Barres de Almeida \inst{8} \thanks{supported by CAPES Foundation, Ministry of Education of Brazil}
 \and A.R.~Bazer-Bachi \inst{3}
 \and B.~Behera \inst{14}
 \and M.~Beilicke \inst{4}
 \and W.~Benbow \inst{1}
 %\and D.~Berge \inst{1} \thanks{now at CERN, Geneva, Switzerland}
 \and K.~Bernl\"ohr \inst{1,5}
 \and C.~Boisson \inst{6}
 %\and O.~Bolz \inst{1}
 \and V.~Borrel \inst{3}
% ...
}

\institute{
Max-Planck-Institut f\"ur Kernphysik, Heidelberg, Germany
\and
 Yerevan Physics Institute, Yerevan, Armenia
\and
Centre d'Etude Spatiale des Rayonnements, CNRS/UPS, Toulouse, France
\and
Universit\"at Hamburg, Institut f\"ur Experimentalphysik, Hamburg, Germany
\and
Institut f\"ur Physik, Humboldt-Universit\"at zu Berlin, Berlin, Germany
\and
LUTH, Observatoire de Paris, CNRS, Universit\'e Paris Diderot, Meudon, France\\
\email{\href{mailto:jean-philippe.lenain@obspm.fr}{jean-philippe.lenain@obspm.fr}}
\and
IRFU/DSM/CEA, CE Saclay, Gif-sur-Yvette, France
\and
University of Durham, Department of Physics, Durham, UK
\and
Unit for Space Physics, North-West University, Potchefstroom, South Africa
\and
Laboratoire Leprince-Ringuet, Ecole Polytechnique, CNRS/IN2P3, Palaiseau, France
\and 
Laboratoire d'Annecy-le-Vieux de Physique des Particules, CNRS/IN2P3, Annecy-le-Vieux, France
\and
Astroparticule et Cosmologie (APC), CNRS, Universite Paris 7 Denis Diderot, Paris, France
\thanks{UMR 7164 (CNRS, Universit\'e Paris VII, CEA, Observatoire de Paris)}
\and
Dublin Institute for Advanced Studies, Dublin, Ireland
\and
Landessternwarte, Universit\"at Heidelberg, Heidelberg, Germany
% ...
}

\offprints{J.-P.~Lenain, D.~Nedbal\\
  \email{\href{mailto:jean-philippe.lenain@obspm.fr}{jean-philippe.lenain@obspm.fr},\\\href{mailto:dalibor.nedbal@mpi-hd.mpg.de}{dalibor.nedbal@mpi-hd.mpg.de}}}

\date{Received 19 February 2008 / Accepted 26 February 2008}

% \abstract{}{}{}{}{} 
% 5 {} token are mandatory

\abstract
    {} %leave it empty if necessary
    {
      The BL\,Lac object \rgb\ ($z=0.080$) was predicted to be a very high-energy (VHE; $> 100$\,GeV) $\gamma$-ray source, due to its high X-ray and radio fluxes. Our aim is to understand the radiative processes by investigating the observed emission and its production mechanism using the High Energy Stereoscopic System (H.E.S.S.) experiment.
    }
    {
      We report recent observations of the BL\,Lac source \rgb\ made in late October and November 2007 with the H.E.S.S. array consisting of four imaging atmospheric Cherenkov telescopes. Contemporaneous observations were made in X-rays by the {\textit{Swift\/}} and {\textit{RXTE\/}} satellites, in the optical band with the ATOM telescope, and in the radio band with the Nan\c{c}ay Radio Telescope.
    }
    {
      A signal of 173 $\gamma$-ray photons corresponding to a statistical significance of 6.6\,$\sigma$ was found in the data. The energy spectrum of the source can be described by a powerlaw with a spectral index of \spectralindex. The integral flux above 300\,GeV corresponds to $\sim$2\% of the flux of the Crab nebula. The source spectral energy distribution (SED) can be described using a two-component non-thermal synchrotron self-Compton (SSC) leptonic model, except in the optical band, which is dominated by a thermal host galaxy component. The parameters that are found are very close to those found in similar SSC studies in TeV blazars.
    }
    {
      \rgb\ is discovered as a source of VHE $\gamma$-rays by H.E.S.S. The location of its synchrotron peak, as derived from the SED in {\textit{Swift\/}} data, allows clear classification as a high-frequency-peaked BL\,Lac (HBL).
    }
    
    
    \maketitle
    %
    %________________________________________________________________
    
    \section{Introduction}
    
    First detected as a radio source \citep{1991ApJS...75....1B} by the NRAO Green Bank Telescope and in the Parkes-MIT-NRAO surveys \citep{1995ApJS...97..347G}, \rgb\ was later identified as a BL\,Lac object by \citet{1998ApJS..118..127L}, who located it at $z=0.080$, and was claimed as an intermediate-frequency-peaked BL\,Lac object by \citet{1999ApJ...525..127L}. \citet{1997A&A...323..739B} report the first detection of \rgb\ in X-rays in the {\it ROSAT}-Green Bank (RGB) sample. The host is an elliptical galaxy with luminosity $M_R=-24.0$ \citep{2003A&A...400...95N}. The source has high radio and X-ray fluxes, making it a good candidate for VHE emission \citep{2002A&A...384...56C}, motivating its observation by the H.E.S.S. experiment.
    
    The broad-band SED of BL\,Lac objects is typically characterised by a double-peaked structure, usually attributed to synchrotron radiation in the radio-to-X-ray domain and inverse Compton scattering in the $\gamma$-ray domain, which is frequently explained by SSC models \citep[see, e.g.,][]{2005A&A...442..895A}. However, since the flux of BL\,Lac objects can be highly variable \citep[e.g.][]{2000A&A...353...97K}, stationary versions of these models are only relevant for contemporaneous multi-wavelength observations of a non-flaring state. The contemporaneous radio, optical, X-ray, and VHE observations presented here do not show any significant variability, and thus enable the first SSC modelling of the emission of \rgb.

    \section{H.E.S.S. observations and results}
    
    %\subsection{Observations}
    
    \rgb\ was observed by the H.E.S.S. array consisting of four imaging atmospheric Cherenkov telescopes, located in the Khomas Highland, Namibia \citep{2006A&A...457..899A}. The observations were performed from October 30 to November 14, 2007. The data were taken in {\em wobble} mode, where the telescopes point in a direction typically at an offset of 0.5\degr\ from the nominal target position \citep{2006A&A...457..899A}. After applying selection cuts to the data to reject periods affected by poor weather conditions and hardware problems, the total live-time used for analysis amounts to 14.7\,h. The mean zenith angle of the observations is $26.9$\degr. 
    
    The data are calibrated according to \citet{2004APh....22..109A}. Energies are reconstructed taking the effective optical efficiency evolution into account \citep{2006A&A...457..899A}. The separation of $\gamma$-ray-like events from cosmic-ray background events was made using the Hillas moment-analysis technique \citep{1985ICRC....3..445H}. Signal extraction was performed using {\em standard cuts} \citep{2006A&A...457..899A}. The on-source events were taken from a circular region around the target with a radius of $\theta=0.11$\degr. The background was estimated using {\em reflected regions} \citep{2006A&A...457..899A} located at the same offset from the centre of the observed field as the on-source region.

    %\subsection{Results}
    
\begin{figure} %fig 1
      \resizebox{\hsize}{!}{\includegraphics{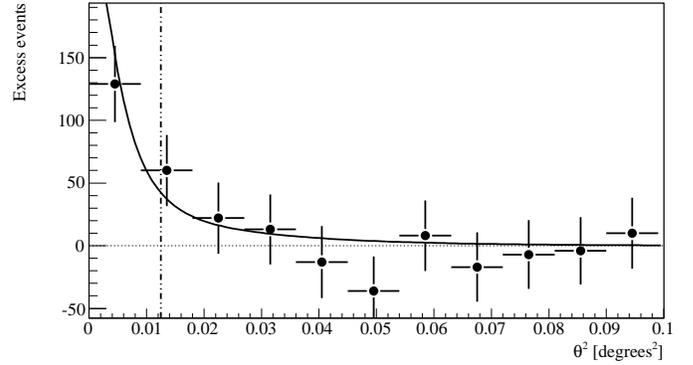}}
      \caption{Angular distribution of excess events. The dot-dashed line shows the angular distance cut used for extracting the signal. The excess distribution is consistent with the H.E.S.S. point spread function as derived from Monte Carlo simulations (solid line).
      }
      \label{fig:thetasq}
    \end{figure}

    A signal of 173 $\gamma$-ray events is found from the direction of \rgb. The statistical significance of the detection is 6.6 $\sigma$ according to \citet{1983ApJ...272..317L}. The preliminary detection was reported by \citet{2007ATel.1295....1N}. A two-dimensional Gaussian fit of the excess yields a position $\alpha_{J2000}=1^{\mathrm h}52^{\mathrm m}33 \fs 5 \pm 5 \fs 3_{\rm stat}\pm 1 \fs 3_{\rm syst}$, $\delta_{J2000}=1^{\circ}46' 40 \farcs 3 \pm 107''_{\rm stat}\pm 20''_{\rm syst}$. The measured position is compatible with the nominal position of \rgb\ ($\alpha_{J2000}=1^{\mathrm h}52^{\mathrm m}39\fs78$, $\delta_{J2000}=1^{\circ}47'18\farcs70$) at the 1$\sigma$ level. Given this spatial coincidence, we identify the source of $\gamma$-rays with \rgb. The angular distribution of events coming from \rgb, shown in Fig.~\ref{fig:thetasq}, is compatible with the expectation from the Monte Carlo simulations of a point source.
    
    %______________________________________________________________
    %
    %
    %______________________________________________________________

    %______________________________________________________________
    \begin{figure} %fig 2
      \resizebox{\hsize}{!}{\includegraphics{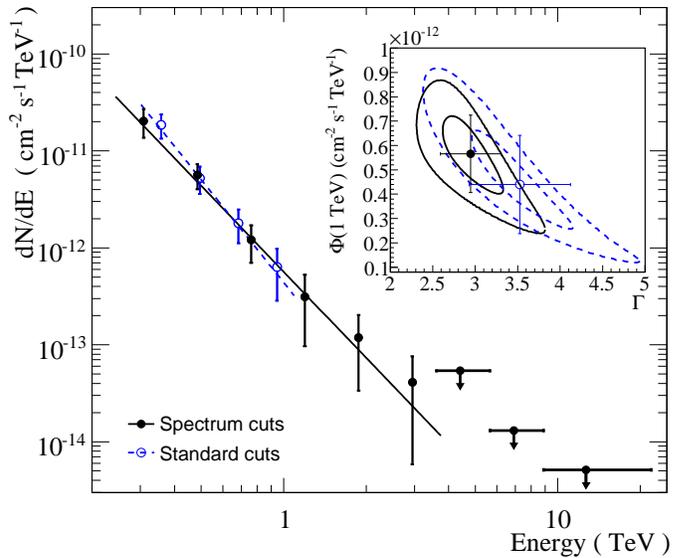}}
      \caption{Differential spectrum of \rgb. The spectrum obtained using {\em spectrum cuts} (black closed circles) is compared with the one obtained by the {\em standard cuts} (blue open circles). The black line shows the best fit by a powerlaw function of the former. The three points with the highest photon energy represent upper limits at 99\% confidence level, calculated using \citet{1998PhRvD..57.3873F}. All error bars are only statistical. The fit parameters of a powerlaw fit are \spectralindex\ and \normalization\ for the {\em spectrum cuts}, and \spectralindexstd\ and \normalizationstd\ for the {\em standard cuts}. The insert shows 1 and 2\,$\sigma$ confidence levels of the fit parameters.
      }
      \label{fig:spectrum}
    \end{figure}
    %
    %
    %______________________________________________________________

    Figure~\ref{fig:spectrum} shows the time-averaged differential spectrum. The spectrum was derived using {\em standard cuts} with an energy threshold of 300\,GeV. Another set of cuts, the {\em spectrum cuts} described in \citet{2006A&A...448L..19A}, were used to lower the energy threshold and improve the photon statistics (factor $\sim$2 increase above the {\em standard cuts}). Both give consistent results (see inlay in Fig.~\ref{fig:spectrum} and caption). Because of the better statistics and energy range, we use the spectrum derived using {\em spectrum cuts} in the following. Between the threshold of 240\,GeV and 3.8\,TeV, this differential spectrum is described well ($\chi^2$/d.o.f.=2.16/4) by a power law ${\rm d}N/{\rm d}E=\Phi_0(E/ 1\mathrm{TeV})^{-\Gamma}$ with a photon index \spectralindex\ and normalisation at 1\,TeV of \normalization. The 99\% confidence level upper limits for the highest three bins shown in Fig.~\ref{fig:spectrum} were calculated using \citet{1998PhRvD..57.3873F}.

    %______________________________________________________________
    \begin{figure} %fig 3
      \resizebox{\hsize}{!}{\includegraphics{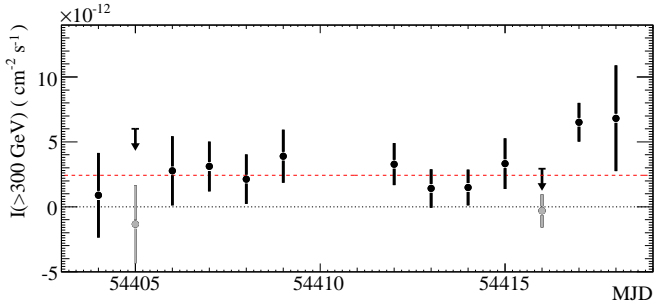}}
      \caption{Mean nightly integral flux from \rgb\ above 300\,GeV. Only the statistical errors are shown. Upper limits at 99\% confidence level are calculated when no signal is found (grey points). The dashed line shows a fit of a constant to the data points with $\chi^2/\rm{d.o.f.}$ of $17.2/12$. The fit was performed using all nights.}
      \label{fig:lightcurve}
    \end{figure}
    %
    %
    %______________________________________________________________

    The integral flux above 300\,GeV is $I = (2.70\pm0.51_{\mathrm{stat}}\pm 0.54_{\mathrm{syst}})\times 10^{-12}$\fluxunits, which corresponds to $\sim$2\% of the flux of the Crab nebula above the same threshold as determined by \citet{2006A&A...457..899A}. Figure~\ref{fig:lightcurve} shows the nightly evolution of the $\gamma$-ray flux above 300\,GeV. There is no significant variability between nights in the lightcurve. The $\chi^2/\rm{d.o.f.}$ of the fit to a constant is $17.2/12$, corresponding to a $\chi^2$ probability of 14\%.
    
    All results were checked with independent analysis procedures and calibration chain giving consistent results.

    \section{Multi-wavelength observations with \textit{Swift}, \textit{RXTE}, ATOM, and the Nan\c{c}ay Radio Telescope}
    
    \subsection{X-ray data from {\it Swift} and {\it RXTE}}
    
    Target of opportunity (ToO) observations of \rgb\ were performed with {\it Swift} and {\it RXTE} on November 13, 14, and 15, 2007 triggered by the H.E.S.S. discovery.
    
    The {\it Swift}/XRT \citep{2005SSRv..120..165B} data (5.44\,ks) were taken in photon-counting mode. The spectra were extracted with {\tt xselect v2.4} from a circular region with a radius of 20 pixels ($0.8\arcmin$) around the position of \rgb, which contains $90\%$ of the PSF at 1.5\,keV. An appropriate background was extracted from a region next to the source with four times this area. The auxiliary response files were created with the script {\tt xrtmkarf v0.5.6} and the response matrices were taken from the {\it Swift} package of the calibration database {\tt caldb v3.4.1}. Due to the low count rate of $0.3\,\rm{cts/s}$, any pileup effect on the spectrum is negligible. We find no significant variability during any of the pointings or between the three subsequent days; hence, individual spectra were combined to achieve better photon statistics. The spectral analysis was performed using the tool  {\tt Xspec v11.3.2}. A broken powerlaw ($\Gamma_1 =1.93 \pm 0.20, \Gamma_2 = 2.82 \pm 0.13, E_\mathrm{break} = 1.29 \pm 0.12 \, \rm{keV}$) with a Galactic absorption of $2.72 \times 10^{20}\,\rm{cm^{-2}}$ \citep[LAB Survey,][]{2005A&A...440..775K} is a good description ($\chi^2 / \rm{d.o.f.} = 24 / 26$), and the resulting unabsorbed flux is $F_{0.5 - 2 \,\rm{keV}}\sim 5.1 \times10^{-12}\,\rm{erg \,cm^{-2}\, s^{-1}}$ and $F_{2 - 10 \,\rm{keV}}\sim 2.7 \times10^{-12}\,\rm{erg \,cm^{-2}\, s^{-1}}$.
    
    Simultaneous observations at higher X-ray photon energies were obtained with the {\it RXTE}/PCA \citep{1996SPIE.2808...59J}. Only data of PCU2 and the top layer were taken to obtain the best signal-to-noise ratio. After filtering out the influence of the South Atlantic Anomaly, tracking offsets, and the electron contamination, an exposure of 3.2\,ks remains. Given the low count rate of $0.7\,\rm{cts/s}$, the ``faint background model'' provided by the {\it RXTE} Guest Observer Facility was used to generate the background spectrum with the script {\tt pcabackest v3.1}. The response matrices were created with {\tt pcarsp v10.1}. Again no significant variations were found between the three observations, and individual spectra were combined to achieve better photon statistics. The PCA spectrum can be described by an absorbed single powerlaw with photon index $\Gamma = 2.72 \pm 0.08$ ($\chi^2 / \rm{d.o.f.} = 20/16$) between 2 and 10\,keV, using the same Galactic absorption as for {\it Swift} data. The resulting flux $F_{2 - 10 \,\rm{keV}}\sim 6.8 \times10^{-12}\,\rm{erg \,cm^{-2}\, s^{-1}}$ exceeds the one obtained {\em simultaneously} with {\it Swift} by a factor of 2.5. We attribute this mostly to contamination by the nearby galaxy cluster Abell\,267 ($44.6\arcmin$ offset from \rgb\ but still in the field of view of the non-imaging PCA).

    A detailed decomposition is beyond the scope of this paper, so we exclude {\it RXTE} data from broadband modelling. The {\it RXTE} data-set confirms the absence of variability during November 2007, also in the energy range up to 10 keV. For the SED modelling, the average spectrum is treated as an upper limit. Further observations with {\it RXTE} in December 2007 also show no indication of variability.

%, and will be presented in a more detailed subsequent publication.

    \subsection{Optical data}
    
    Optical observations were taken using the ATOM telescope \citep{2004AN....325..659H} at the H.E.S.S site from November 10, 2007. No significant variability was detected during the nights between November 10 and November 20; R-band fluxes binned nightly show an RMS of 0.02\,mag.
    
    Absolute flux values were found using differential photometry against stars calibrated by K.\,Nilsson (priv.~comm.). We measured a total flux of $m_R = 15.25 \pm 0.01$\,mag (host galaxy + core) in an aperture of $4\arcsec$ radius. The host galaxy was subtracted with galaxy parameters given in \citet{2003A&A...400...95N}, and aperture correction given in Eq.~(4) of \citet{1976AJ.....81..807Y}. The core flux in the R-band (640\,nm) was found to be 0.62 $\pm$ 0.08\,mJy. This  value was not corrected for Galactic extinction.

    \subsection{Radio data}
    
    The Nan\c{c}ay Radio Telescope (NRT) is a meridian transit telescope with a main spherical mirror of 300\,m $\times$ 35\,m \citep{2007A&A...465...71T}. The low-frequency receiver, covering the band 1.8--3.5\,GHz was used, with the NRT standard filterbank backend.
    
    The NRT observations were obtained in two contiguous bands of 12.5\,MHz bandwidth, centred at 2679 and 2691\,MHz (average frequency: 2685\,MHz). Two linear polarisation receivers were used during the 22 60-second drift scan observations on the source on November 12 and 14, 2007. The data have been processed with the standard NRT software packages NAPS and SIR. All bands and polarisations have been averaged, giving an RMS noise of 2.2\,mJy. The source 3C\,295 was observed for calibration, on November 11, 13, and 15, 2007.
    
    Taking into account a flux density for this source of $12.30 \pm 0.06$\,Jy using the spectral fit published by \citet{1994A&A...284..331O}, we derived a flux density of $56 \pm 6$\,mJy at 2685\,MHz for \rgb. No significant variability was found in the radio data.

    \section{Discussion}
    \label{sec:modeling}

    %______________________________________________________________
    \begin{figure*}[tbh]  %fig 4
      \centering
      \begin{minipage}[c]{0.65\textwidth}
        \includegraphics[angle=-90,width=\textwidth]{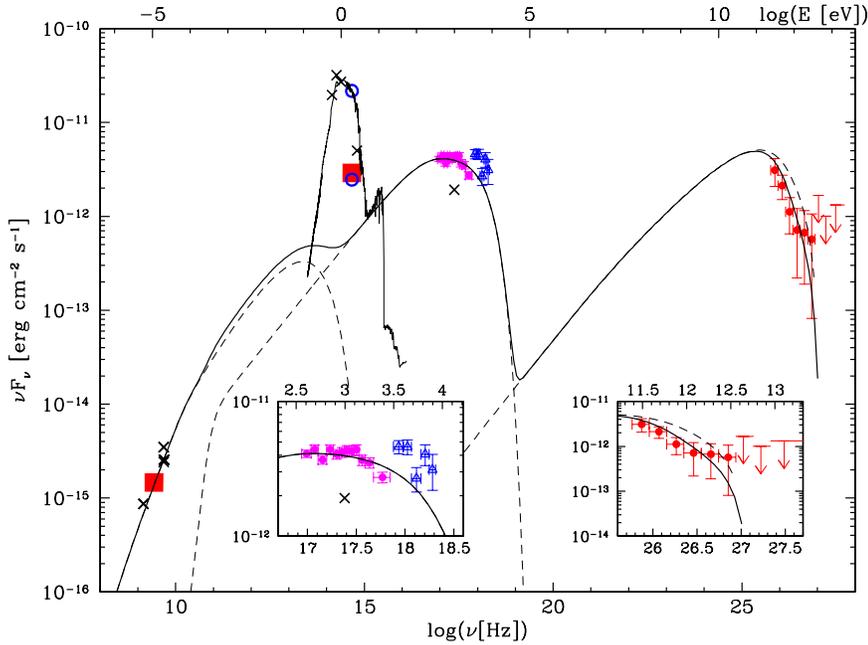}
      \end{minipage}%
      \begin{minipage}[c]{0.35\textwidth}
        \caption{The spectral energy distribution of \rgb. Shown are the H.E.S.S. spectrum ({\it red filled circles and upper limits\/}), and contemporaneous {\it RXTE} ({\it blue open triangles\/}), {\it Swift}/XRT (corrected for Galactic absorption, {\it magenta filled circles}), optical host galaxy-subtracted (ATOM) and radio (Nan\c{c}ay) observations ({\it large red filled squares\/}). The black crosses are archival data. The blue open points in the optical R-band correspond to the total and the core fluxes from \citet{2003A&A...400...95N}. A blob-in-jet synchrotron self-Compton model (see text) applied to \rgb\ is also shown, describing the soft X-ray and VHE parts of the SED, with a simple synchrotron model shown at low frequencies to describe the extended part of the jet. The contribution of the dominating host galaxy is shown in the optical band. The dashed line above the solid line at VHE shows the source spectrum after correcting for EBL absorption. The left- and right-hand side inlays detail portions of the observed X-ray and VHE spectrum, respectively.
        }
      \end{minipage}%
      \label{fig:SED}
    \end{figure*}
    %
    %
    %______________________________________________________________
    
    Figure~\ref{fig:SED} shows the SED of \rgb\ with the data from Nan\c{c}ay, ATOM, {\it Swift}/XRT, {\it RXTE}/PCA, and H.E.S.S. Even though some data are not strictly simultaneous, no significant variability is found in the X-ray and optical bands throughout the periods covered; hence, a common modelling of the contemporaneous X-ray and VHE data appears justified.
    
    The optical part of the SED is mainly due to the host galaxy, which is detected and resolved in optical wavelengths \citep{2003A&A...400...95N}. A template of the spectrum of such an elliptic galaxy is shown in the SED, as inferred from the code PEGASE \citep{1997A&A...326..950F}. The host-galaxy-subtracted data point from the ATOM telescope might include several additional contributions---from an accretion disk, an extended jet (see below), or a central stellar population---so that it is considered as an upper limit in the following SSC model. A model including the optical ATOM data with possible additional contributions is beyond the scope of this paper.

%, and will be treated in a subsequent publication.

    We applied a non-thermal leptonic SSC model \citep{2001A&A...367..809K} to account for the contemporaneous observations by {\it Swift} in X-rays and by H.E.S.S. in the VHE band. The radio data are assumed to originate in an extended region, described by a separate synchrotron model for the extended jet \citep{2001A&A...367..809K} to explain the low-frequency part of the SED \citep[as in, e.g.,][]{2005A&A...442..895A,2008A&A...477..481A}.

    We should emphasise that the aim of applying this model in this work is not to present a definitive interpretation for this source, but rather to show that a standard SSC model is able to account for the VHE and {\it Swift} X-ray observations.
    
    For the SSC model, we describe the system as a small homogeneous spherical, emitting region (blob) of radius $R$ within the jet, filled with a tangled magnetic field $B$ and propagating with a Doppler factor $\delta=\left[ \Gamma \left( 1 -\beta \cos{\theta} \right) \right]^{-1}$. Here $\Gamma$ is the bulk Lorentz factor of the emitting plasma blob, $\beta = v/c$, and $\theta$ is the angle of the velocity vector, with respect to the line-of-sight. The electron energy distribution (EED) is described by a broken powerlaw, with indices $n_1$ and $n_2$, between Lorentz factors $\gamma_\mathrm{min}$ and $\gamma_\mathrm{max}$, with a break at $\gamma_\mathrm{break}$ and density normalisation $K$.
    
    The model also accounts for the absorption by the extragalactic background light (EBL) with the parameters given in \citet{2005AIPC..745...23P}. \rgb\ is too nearby ($z=0.08$) to add to the constraints on the EBL that were found by H.E.S.S. measurements of other blazars \citep{2006Natur.440.1018A}. In all the models, we assume $H_0 = 70$\,km\,s$^{-1}$\,Mpc$^{-1}$, giving a luminosity distance of $d_L = 1.078 \times 10^{27}$\,cm for \rgb.
    
    The EED can be described by $K=3.1~\times~10^4$\,cm$^{-3}$, $\gamma_\mathrm{min}=1$, and $\gamma_\mathrm{max}=4~\times~10^5$. The break energy is assumed at $\gamma_\mathrm{break}=7.0~\times~10^4$ and is consistent with the {\it Swift}/XRT spectrum, while providing good agreement with the H.E.S.S. data. We assume the canonical index $n_1=2.0$ for the low-energy part of the EED, in accordance with standard Fermi-type acceleration mechanisms. The value $n_2=3.0$ for the high-energy part of the EED is constrained by the high-energy part of the X-ray spectrum. A good solution is found with the emitting region characterised by $\delta=25$, $R=1.5~\times~10^{15}$\,cm, and $B=0.10$\,G.
    
    For the extended jet, the data are described well by $R_\mathrm{jet}=10^{16}$\,cm, $\delta_\mathrm{jet}=7$, $K_\mathrm{jet}=70$\,cm$^{-3}$, $B_\mathrm{jet}=0.05$\,G, and $\gamma_\mathrm{break,\;jet}=10^4$ at the base of the jet, and $L_\mathrm{jet}=50$\,pc \citep[all the parameters are detailed in][]{2001A&A...367..809K}.
    
    Assuming additional contributions in the optical band, the multi-wavelength SED can thus be explained with a standard shock-acceleration process. The parameters derived from the model are similar to previous results for this type of source \citep[see, e.g.,][]{2002A&A...386..833G}.

    From the current Nan\c{c}ay radio data and the {\it Swift} X-ray data, we obtain a broad-band spectral index $\alpha_{rx} \sim 0.56$ between the radio and the X-ray domains. The obtained SED, the corresponding location of the synchrotron peak, and the flux and shape of the {\it Swift} spectrum lead us to conclude that \rgb\ can clearly be classified as an HBL object at the time of H.E.S.S. observations.

    \section{Conclusion}
    
    The HBL \rgb\ was detected in VHE at energies $>~300$\,GeV with the H.E.S.S. experiment. The contemporaneous {\it Swift}, {\it RXTE}, Nan\c{c}ay, ATOM, and H.E.S.S. data allow the multi-wavelength SED for \rgb\ to be derived for the first time , and to clearly confirm its HBL nature at the time of the H.E.S.S. observations. In general, large variations of the VHE flux are expected in TeV blazars, making further monitoring of this source to detect high states of the VHE flux (flares) desirable.

    \begin{acknowledgements}
      The support of the Namibian authorities and of the University of Namibia in facilitating the construction and operation of H.E.S.S. is gratefully acknowledged, as is the support by the German Ministry for Education and Research (BMBF), the Max Planck Society, the French Ministry for Research, the CNRS-IN2P3 and the Astroparticle Interdisciplinary Programme of the CNRS, the U.K. Science and Technology Facilities Council (STFC), the IPNP of the Charles University, the Polish Ministry of Science and Higher Education, the South African Department of Science and Technology and National Research Foundation, and by the University of Namibia. We appreciate the excellent work of the technical support staff in Berlin, Durham, Hamburg, Heidelberg, Palaiseau, Paris, Saclay, and in Namibia in the construction and operation of the equipment.
      
      This research made use of the NASA/IPAC Extragalactic Database (NED). The authors thank the {\it RXTE} team for their prompt response to our ToO request and the professional interactions that followed. The authors acknowledge the use of the publicly available {\it Swift} data, as well as the public HEASARC software packages. This work uses data obtained at the Nan\c{c}ay Radio Telescope. The authors also thank Dr.~Mira V\'eron-Cetty for fruitful discussions.
    \end{acknowledgements}
    
    \bibliographystyle{aa}
    \bibliography{HESS_RGB0152}
    
\end{document}